\documentclass[pra,showpacs,twocolumn]{revtex4}

\usepackage{amsmath}
\usepackage{amssymb}
\usepackage{graphicx}
\usepackage{epsfig}

\begin{document}

\title{Local Operations in qubit arrays via global but periodic Manipulation}
\author{Zheng-Wei Zhou,Yong-Jian Han, and Guang-Can Guo}

\address{Laboratory of Quantum Information, University of Science
and Technology of China, Hefei, Anhui 230026, P. R. China}

\begin{abstract}
We provide a scheme for quantum computation in lattice systems via
global but periodic manipulation, in which only effective periodic
magnetic fields and global nearest neighbor interaction are
required. All operations in our scheme are attainable in optical
lattice or solid state systems. We also investigate universal
quantum operations and quantum simulation in 2 dimensional
lattice. We find global manipulations are superior in simulating
some nontrivial many body Hamiltonians.
\end{abstract}

\maketitle

\section{Introduction}

Since a quantum computer (QC) will exhibit advantages over its
classical counterpart only when a large number of qubits can be
manipulated coherently, many architectures of QCs based on
scalable physical systems have been
investigated\cite{SQC1,SQC2,SQC3,SQC4,SQC5}. Among these
candidates, ultra-cold neutral atoms trapped in the periodic
potentials of an optical lattice are attracting much
attention\cite
{SQC5,SQC6,SQC7,SQC8,Cold-atom1,Cold-atom2,Cold-atom3,Cold-atom4}.
Since neutral atoms couple weakly to the environment, decoherence
is suppressed greatly. Besides, some typical many-body models can
be constructed in optical lattice systems. This makes it possible
to use quantum optical methods to study fundamental
condensed-matter physics problems\cite {Cold-atom5,
Cold-atom6,Cold-atom7,Cold-atom8}.

In the schemes for QCs based on optical lattice system, the
couplings between nearest neighbor atoms in the lattice can be
simultaneously switched on and off by adjusting the intensity,
frequency, and polarization of the trapping light. However, it is
difficult to focus a laser beam on a single atom due to the short
lattice period. This makes it challenging to realize controlled
collisions and perform single and two qubit logical operations. A
possible approach to overcome the difficulty of addressing the
atoms individually is based on marker qubits\cite{SQC6,SQC7,SQC8},
i.e., by using two types of atoms with different internal states
in the optical lattice. One type of atoms act as marker qubits to
address logical qubits. Another method is one way quantum
computation\cite{oneway}. In this proposal, the couplings between
adjacent atoms can be utilized to prepare the atoms in a high
dimensional cluster state. After the initialization, one can
enlarge the lattice period to single qubit addressable range.
Universal quantum computation can then be implemented simply via a
series of single qubit measurements. Of course, one way quantum
computation requires a great qubit overhead.

Recently, a series of schemes for quantum computation based on
global operations only were proposed. In these schemes, local
addressability become
unnecessary\cite{Lloyd,Benjamin,Wu,Raussendorf,Cirac,Nagy}.
Remarkably, Raussendorf devised a novel scheme for universal
quantum computation via translation-invariant operations on a
chain of qubits\cite{Raussendorf}, in which only the nearest
neighbor Ising-type interaction and translation-invariant single
qubit unitary operations are required. Inspired by these
proposals, in this letter we present a scheme for quantum
computation based on global operations in qubit arrays only. Here,
we develop a general method to realize single qubit unitary
operations via global but periodic manipulation. Compared with the
quantum cellular automata scheme\cite {Lloyd,Benjamin}, encoding
overhead is eliminated in our proposal. In addition, single qubit
operation is easier than in Raussendorf's
scheme\cite{Raussendorf}.

The paper is organized as follows. In Sec.II, we show that
universal quantum computation in one dimensional arrays can be
realized via global but periodic manipulation. We also display
that the preparation for initial states and the measurement for
final states can be implemented by using this global manipulation.
Furthermore, in Sec.III, we discuss quantum computation and
quantum simulation in high dimensional arrays. We find the
implementation of the periodic operations in the finite lattice
will benefit from global manipulation. Section IV contains some
conclusions.

\section{Universal quantum computation in one dimensional
arrays}

Let us first consider a two-component bosonic atomic mixture
trapped in one dimension by two spin-dependent lattice potentials.
Each atom is assumed to have two relevant internal states, which
are denoted with the effective spin index $\sigma =\uparrow $,
$\downarrow $, respectively. In the Mott insulator regime, the
system can be described by a two-component Bose-Hubbard
Hamiltonian\cite{SQC5}:
\begin{eqnarray}
H &=&-\sum_{i,\sigma }(t_\sigma a_{i\sigma }^{+}a_{i+1\sigma }+H.c.)+\frac
12\sum_{i,\sigma }U_\sigma n_{i\sigma }(n_{i\sigma }-1)  \nonumber \\
&&+U_{\uparrow \downarrow }\sum_in_{i\uparrow }n_{i\downarrow }.  \label{eq1}
\end{eqnarray}
Here, $a_{i\sigma }^{+}$ and $a_{i\sigma }$ are creation and
annihilation operators for bosonic atoms of spin $\sigma $
localized on-site $i$, and $n_{i\sigma }=a_{i\sigma
}^{+}a_{i\sigma }$. Under the condition that $t_\sigma \ll
U_\sigma ,U_{\uparrow \downarrow }$ and $\left\langle n_{i\uparrow
}\right\rangle +\left\langle n_{i\downarrow }\right\rangle = 1$,
Eq. (1) is equivalent to the following $XXZ$ type interaction
Hamiltonian\cite{SQC5,Kuklov}:
\begin{equation}
H_S=\sum_i\left[ J_1\sigma _i^z\sigma _{i+1}^z-J_2(\sigma _i^x\sigma
_{i+1}^x+\sigma _i^y\sigma _{i+1}^y)\right] .  \label{eq2}
\end{equation}
Here, $\sigma _i^z=n_{i\uparrow }-n_{i\downarrow }$, $\sigma
_i^x=a_{i\uparrow }^{+}a_{i\downarrow }+a_{i\downarrow
}^{+}a_{i\uparrow }$, and $\sigma _i^y=-i(a_{i\uparrow
}^{+}a_{i\downarrow }-a_{i\downarrow }^{+}a_{i\uparrow })$ satisfy
$\left[ \sigma _i^\alpha ,\sigma _i^\beta \right] =2i\epsilon
_{\alpha \beta \gamma }\sigma _i^\gamma $. $ J_1=(t_{\uparrow
}^2+t_{\downarrow }^2)/2U_{\uparrow \downarrow }-t_{\uparrow
}^2/U_{\uparrow }-t_{\downarrow }^2/U_{\downarrow }$, $
J_2=t_{\uparrow }t_{\downarrow }/U_{\uparrow \downarrow }$.
Remarkably, the parameters $J_1$ and $J_2$ can be easily
controlled by adjusting the intensity of the trapping laser beam
or an external field\cite{SQC5}. Therefore, the following
well-known spin $1/2$ interaction models can be realized: Ising
model $\left( J_1\neq 0,J_2=0\right) $, $XY$ model $\left(
J_1=0,J_2\neq 0\right) $, and Heisenberg antiferromagnetic
(ferromagnetic) model $\left( J_1=\pm J_2\right) $.

Besides Hamiltonian $H_S$, in order to implement universal quantum
computation, we introduce the following effective periodic
magnetic field:
\begin{equation}
H_B^\alpha (a,\varphi )=\sum_iA\left[ 1-\cos \left( \frac{2\pi i
}a+\varphi \right) \right] \sigma _i^\alpha ,(\alpha =x,y,z)
\label{eq3}
\end{equation}
Here, we set the distance between the nearest neighbor sites as
unity. $a$ is the period of the magnetic field, when $a\rightarrow
\infty $, $H_B^\alpha (a,\varphi )$ reduces to a homogeneous
magnetic field. To realize this periodic magnetic field, we set
the eigenstates $\left| \uparrow \right\rangle $ and $\left|
\downarrow \right\rangle $ to have the same energy. By the left
and the right circularly polarized light, they are separately
coupled to the common exited level $\left| e\right\rangle $ with a
blue detuning $\Delta $. In such a 3-level system, we can obtain
the effective Hamiltonians $\sigma _x$, $\sigma _y$, and $\sigma
_z$ in the 2 dimensional Hilbert space spanned by $\left| \uparrow
\right\rangle $ and $\left| \downarrow \right\rangle $ only by
adjusting the polarization of coupling light\cite {SQC5}. Thus,
for the cold atoms trapped in one dimensional lattice, we may
construct such a periodic potential field $H_B^\alpha (a,\varphi
)$ along the lattice direction by applying a monochromatic
standing wave laser beam. In addition, in our scheme, we require
the ability to adjust the period of the effective magnetic field
$H_B^\alpha (a,\varphi )$. To this end, we split a beam of
monochromatic light into two beams and make them interfere. Thus,
we can obtain a one dimensional standing wave along the interior
bisector of the angle between two beams of lights, whose period
depends on the angle between the two beams of lights.

\subsection{Single qubit operations}

Here, we present a general method to control any single qubit on
one-dimensional finite lattice by using global Hamiltonian
$H_B^\alpha (a,\varphi )$. Our idea for local manipulation derives
from Fourier transformation i.e., any spatial function can be
decomposed into a superposition of periodic eigen-functions. Let
us consider a one-dimensional chain of $N$ qubits. The unitary
evolution controlled by the effective periodic magnetic filed
$H_B^\alpha (a,\varphi )$ is: $U_1(\theta ,a,k,\alpha )=\exp
\left\{ -i\frac \theta AH_B^\alpha (a,\frac{-2\pi k}a)\right\} $.
Our aim is to implement a delta function type of operation on the
finite spin chain by assembling these wave-like unitary operations
$U_1(\theta ,a,k,\alpha )$. Without loss of generality, we
consider performing a unitary operation $\exp (-i\theta \sigma
_k^x)$ on the $k$th site. The operation on a single site can be
realized in an iterative way. In the iterative procedure, a key
step is to construct parity unitary operation: $F_k^{(1)}=U_1(\pi
/4 ,2,k,z)F_k^{(0)}U_1^{*}(\pi /4,2,k,z)F_k^{(0)}=\prod_{\lceil
\frac{k-1}2\rceil \leq j\leq \lceil \frac{N-k}2\rceil }\exp
(-i2\theta \sigma_{k+2j}^x).$ Here, $\lceil \eta \rceil $ refers
to the integer part of the number $\eta $ and $F_k^{(0)}=\exp
\left\{ -i\frac \theta {2A}H_B^x(\infty ,\pi )\right\}$. The
parity unitary operation $F_k^{(1)}$ only remains homogeneous on
even sites array from the $k$th site. It effectively just
implements a homogeneous operation on a spin chain of $N/2$
qubits. We find that similar parity-eliminated operations can be
established iteratively: $F_k^{(l)}=U_1(\pi /4
,2^l,k,z)F_k^{(l-1)}U_1^{*}(\pi /4,2^l,k,z)F_k^{(l-1)},$ $1\leq
l\leq m,m=\lceil \log _2\left( N-k\right) \rceil +1.$ Finally,
operation on the $k$th qubit can be realized: $F_k^{(m)}=$ $\exp
(-i2^m\theta \sigma _k^x)$ (see Fig. 1(a)). In this iterative
procedure, implementing single qubit rotation takes $3\times
2^m-2$ elementary steps (applications of $U_1(\theta ,a,k,\alpha
)$), which roughly ranges from $1.5N$ to $6N$.

\begin{figure}[tbh]
\epsfig{file=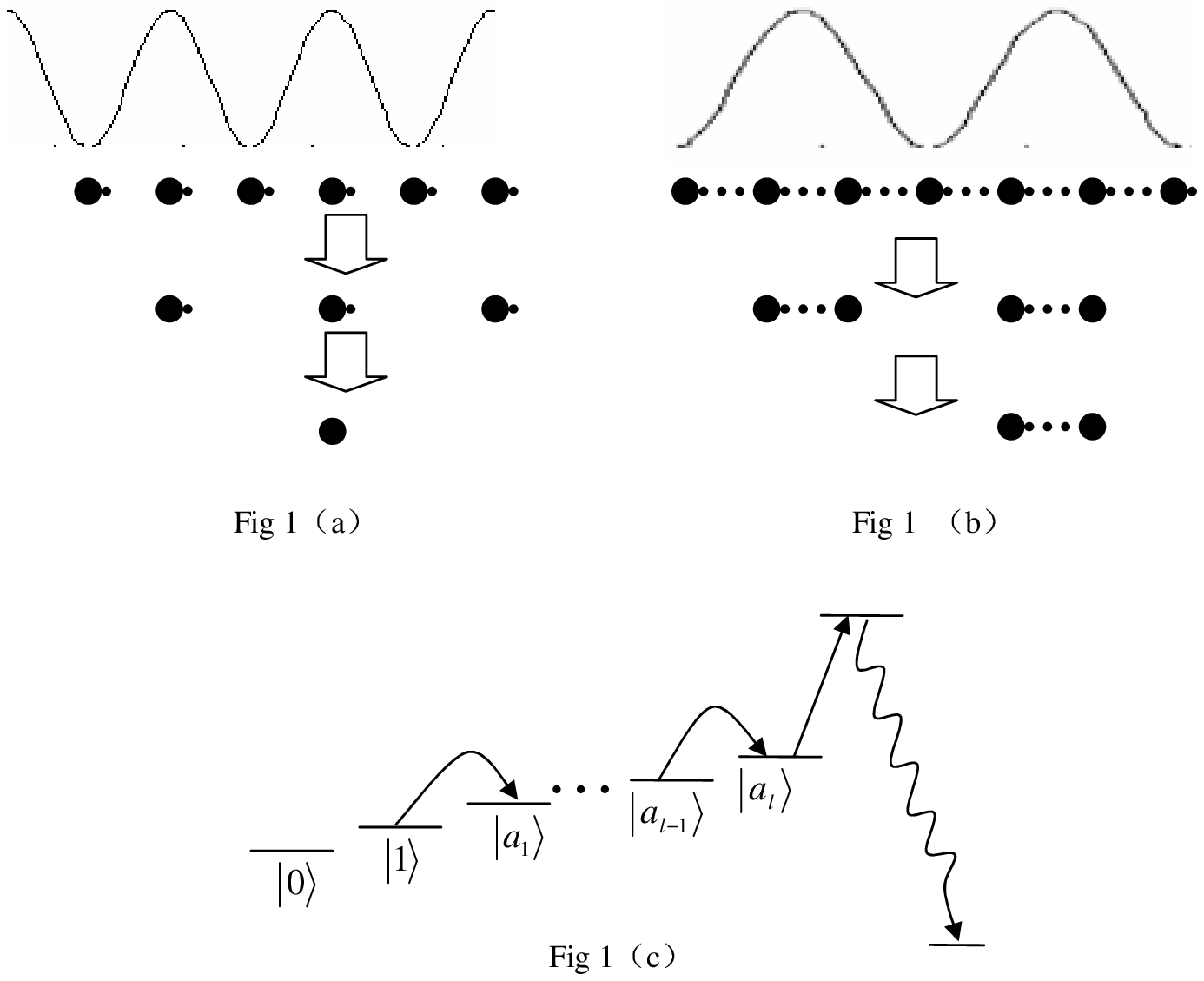,width=8cm} \caption{ Schematic for
addressing and readout measurements via global operations. (a) The
process of single qubit operation. In an iterative step,
approximately half sites are selected out. (b) A high efficiency
procedure for operation between two nearest neighbor sites. In an
iterative step, approximately two thirds couplings in the
remaining spin chain are eliminated. (c) Mapping quantum states
$\left| 1\right\rangle $ localized in far apart sites into
auxiliary energy level $\left| a_l\right\rangle $ by using
modified periodic magnetic fields. The read out for state $\left|
a_l\right\rangle $ can be implemented by the method of detecting
the fluorescence. }
\end{figure}

\subsection{Two-qubit operations}

Since single qubit operations can be realized, implementing
two-qubit operation between two nearest neighbor qubits via many
body interaction becomes feasible. We consider the case of Ising
model: $H_{I\sin g}=\sum_{i=1}^{N-1}J_1\sigma _i^z\sigma
_{i+1}^z$. When the couplings between nearest neighbor qubits
switch on, the corresponding unitary evolution is $U_I(\theta
=J_1t)=\prod_i\exp (i\theta \sigma_i^z\sigma _{i+1}^z)$. To
implement the interaction between the $k$th and $ (k+1)$th qubits
we may synthesize the following unitary transformations: $\exp
(i\theta \sigma _k^z\sigma _{k+1}^z)=U_I(\frac \theta 4)\sigma
_k^xU_I(-\frac \theta 4)\sigma _k^x\sigma _{k+1}^xU_I(-\frac
\theta 4)\sigma _k^xU_I(\frac \theta 4)\sigma _k^x\sigma
_{k+1}^x.$

In the above manipulation, a two-qubit operation takes 6 single
qubit operations, so $9N\sim 36N$ elementary steps (applications
of $ U_1(\vartheta ,a,k,\alpha )$ or $U_I(\vartheta )$) are
required. This overhead can be greatly reduced by devising a new
focusing procedure, which is shown in Fig. 1(b). This focusing
procedure is a bit similar to that of single qubit operation. The
first iterative step is: $T_k^{(1)}(\theta )=U_1(\frac \pi
3,3,k-1,x)U_I(\theta )U_1^{*}(\frac \pi 3,3,k-1,x)U_I(\theta
)=\cdot \cdot \cdot \exp (i2\theta \sigma _{k-3}^z\sigma
_{k-2}^z)\exp (i2\theta \sigma _k^z\sigma _{k+1}^z)\exp (i2\theta
\sigma _{k+3}^z\sigma _{k+4}^z)\cdot \cdot \cdot .$ Here, the
script $k$ refers to focusing the interaction between the $k$th
and $(k+1)$th qubits. After the first step, almost two thirds
couplings in this Ising chain are cut off. Then, we construct the
iterative procedure: $T_k^{(n)}(\theta
)=U_1(a_n,3^n,k-\frac{3^n-1}2,x)T_k^{(n-1)}(\theta
)U_1^{*}(a_n,3^n,k-\frac{3^n-1}2,x)T_k^{(n-1)}(\theta ).$ Here,
$a_n$ is determined by the following restrictions:
\begin{eqnarray}
a_n\left[ \cos \left( \frac{3^{n-1}+1}{3^n}\pi \right) -\cos
\left( \frac{3^{n-1}-1}{3^n}\pi \right) \right] &=&\frac{2p+1}2\pi ,  \nonumber \\
a_n\left[ \cos \left( \frac{3^n-1}{3^n}\pi \right) -\cos \left(
\frac{3^{n-1}-1}{3^n}\pi \right) \right] &=&\frac l2\pi ,  \label{eq4} \\
a_n\left[ 1-\cos \left( \frac{3^n-1}{3^n}\pi \right) \right] &=&\frac m2\pi ,
\nonumber
\end{eqnarray}
where $p,l,$ and $m$ are arbitrary integers. Obviously, for any
given precision, we can always find an $a_n$ to satisfy Eq. (4).
Finally, after repeating the iterative steps $n_f\left( \sim
O(\log _3N)\right) $ times, which corresponds to $3\times
2^{n_f}-2$ elementary steps, we obtain our anticipated operation:
$\exp (i2^{n_f}\theta \sigma _k^z\sigma _{k+1}^z)$.

\subsection{Initialization and read out measurements}

To prepare the qubit trapped in each lattice site to $\left|
0\right\rangle $, we drive the system to Mott insulator regime
with one atom occupancy per lattice site. Furthermore, we adjust
the interaction Hamiltonian $H_S$ to Ising type. When $J_1<0$,
Ising type Hamiltonian have twofold degenerate ground states
$\left| 00,...,0\right\rangle $ and $\left| 11,...,1\right\rangle
$. By applying a homogeneous magnetic field $\sum_if\sigma _i^z$
$\left( f<0\right) $, the degeneracy of ground state will be
broken. Eventually, the system will relax to ground state $\left|
00,...,0\right\rangle $ if the environment is cold enough.

To realize read out measurements for the final state of quantum
computation, we may take advantage of auxiliary energy levels of
the atoms (see Fig.1(c)). Our strategy is to transfer internal
state $\left| 1\right\rangle $ to auxiliary energy level $\left|
a_l\right\rangle $ only for qubits localized in far apart sites,
which can be achieved by using modified focusing operations.
Therefore, we may address and read out internal states $\left|
a_l\right\rangle $ in these lattice sites.  Compared with single
qubit operation, we replace periodic magnetic field $\sum_iA\left[
1-\cos \left( \frac{2\pi i}{2^k}+ \varphi \right) \right] \sigma
_i^\alpha $ with $\sum_iA\left[ 1-\cos \left( \frac{2\pi
i}{2^k}+\varphi \right) \right] \sigma _{ki}^\alpha $, where
$\sigma _{ki}^\alpha $ is Pauli operator in Hilbert space spanned
by $\left\{ \left| a_{k-1}\right\rangle _i,\left| a_k\right
\rangle _i\right\} $ (we set $\left| a_0\right\rangle _i=\left|
1\right\rangle _i$). Thus, we may map $\left| 1\right\rangle $ to
$\left| a_1\right\rangle $ for atoms in next nearest neighbor
lattice sites after the first iterative step. Using similar
iterative procedure, we may transfer $\left| 1\right\rangle $ to
$\left| a_l\right\rangle $ only for atoms $2^l$ sites apart. For
the internal state $\left| a_l\right\rangle $, we may address and
read out by the method of detecting the fluorescence.

\section{Quantum computation and quantum simulation in high
dimensional arrays}

Since focusing any single qubit operation and Ising interaction
between any two nearest neighbor qubits can be realized, universal
quantum computation can be implemented in one dimensional qubit
array via global operations. In the described scheme, a quantum
computation on $N$ qubits roughly takes $O(N)$ (exactly less than
$6N$) elementary operations compared with addressable quantum
computation schemes. Although an exponentially increasing number
of steps can be avoided, a linear dependence on the number of
qubits is still a great obstacle for large-scale quantum
computation due to the fragile many body coherence. To reduce the
required resources further we may consider quantum computation in
high dimensional periodic qubit arrays.

Actually, focusing processes in high dimensional lattice are quite
easy because they can be realized just by using one dimensional
focusing methods, repetitively. Here, we just investigate quantum
computation on two dimensional square lattice. In principle,
higher dimensional quantum computation can be easily realized in a
similar way. We set the number of sites in the row and column to
$n$, Thus, the total number of sites in this lattice is $n^2$. For
simplicity, we denote the site of the $i$th row and $j$th column
by $\left( i,j\right) $. First, let us consider a single site on
two dimensional lattice. Without loss of generality, we give the
unitary operation $\exp (-i\theta \sigma _{(i,j)}^x)$ on site
$(i,j)$: $X(i,-\frac \theta 2,x)Y(j,\frac \pi 2,z)X(i,\frac \theta
2,x)Y(j,-\frac \pi 2,z)$, where $ X(i,\frac \theta
2,x)=\prod_{k=1}^n\exp (-i\frac \theta 2\sigma _{(i,k)}^x)$ and
$Y(j,\frac \pi 2,z)=i^n\prod_{k=1}^n\sigma _{(k,j)}^z$. In this
process, only $O(n)$ elementary steps are needed. Next, we
consider implementing Ising interaction between sites $\left(
i,j\right)$ and $\left( i,j+1\right) $. Since focusing operation
on a single site is attainable, a straightforward way to realize
the operation $\exp (i\theta \sigma _{(i,j)}^z\sigma
_{(i,j+1)}^z)$ is: $U_I^{\prime }(\frac \theta 4)\sigma
_{(i,j)}^xU_I^{\prime }(-\frac \theta 4)\sigma
_{(i,j)}^xU_I^{\prime }(\frac \theta 4)\sigma
_{(i,j+1)}^xU_I^{\prime }(-\frac \theta 4)$ $\sigma
_{(i,j+1)}^xU_I^{\prime }(-\frac \theta 4)\sigma _{(i,j)}^x\sigma
_{(i,j+1)}^xU_I^{\prime }(\frac \theta 4)\sigma _{(i,j)}^x\sigma
_{(i,j+1)}^x.$ Here, $U_I^{\prime }(\theta )=\exp \{-i\theta
\sum_{k\leq l}[\sigma _{(k,l)}^z\sigma _{(k\pm 1,l)}^z+\sigma
_{(k,l)}^z\sigma _{(k,l\pm 1)}^z]\}$ . In this focusing procedure,
8 operations on single sites are required. To reduce the
operational resources, we can also consider the simplified scheme.
As shown by the above section, we can use the more efficient
proposal to implement the interactions between the nearest
neighbor sites in one dimension, i.e. to produce an operation $
\exp \{-i\theta \sum_k\sigma _{(k,j)}^z\sigma _{(k,j+1)}^z\}$ via
$O(\log _3n)$ iterative steps. Hence, we just need $2$ single site
operations to achieve our goal: $\exp (i\theta \sigma
_{(i,j)}^z\sigma _{(i,j+1)}^z)=\exp \{i\frac \theta 2\sum_k\sigma
_{(k,j)}^z\sigma _{(k,j+1)}^z\}\sigma _{(i,j)}^z$ $\exp \{-i\frac
\theta 2\sum_k\sigma _{(k,j)}^z\sigma _{(k,j+1)}^z\}\sigma
_{(i,j)}^z.$

From the above proposal, we may find that the quantity of
operations will be saved further for quantum computation in high
dimensional qubits arrays. Similar analysis will show that,
compared with addressable quantum computation schemes,
$O(^d\sqrt{N})$ times numbers of operations are required for the
scheme in a $d$-dimensional qubits array via global operations.
Here, $N$ refers to the total number of the qubits. Then, we may
make a rough comparison for local addressable and global
addressable schemes from the viewpoint of the error of the unitary
evolution. We set the single step average error rate $\varepsilon
$ for local operations and $\xi $ for global ones. For simplicity,
we look on $\varepsilon $ and $\xi $ as the upper bounds of two
types of single step errors. As far as the scheme via global
operations is concerned, an overhead of $O(^d\sqrt{N})$ times
numbers of operations is cost compared with addressable ones.
While, Ref\cite {Nielsen} shows that the upper bound of the error
of the unitary evolution will linearly increase with the number of
unitary transformations. Therefore, if $\varepsilon
>O(^d\sqrt{N})\xi $ one can benifit from the scheme via global
operations. While, in the above analysis, we do not differentiate
the upper bounds for the different types of operations, which will
lead to more subtle comparison for the overhead of operations in
some well-known algorithms. A further analysis for the upper bound
of the error is necessary and meaningful. In addition, an
interesting question remains open: what is the best strategy for
quantum computation via global operations?

Once universal quantum computation can be implemented, any
$SU\left( 2^n\right) $ unitary transformation can be achieved.
Therefore, in principle, quantum computers have the ability to
simulate dynamical behavior of any finite dimensional quantum
systems. However, it is a formidable task to decompose the
time-dependent unitary evolution under some non-trivial many body
Hamiltonians into a sequence of elementary quantum gate
operations. Fortunately, one can use Trotter formula to
approximately implement quantum simulation:
\begin{equation}
e^{-i\sum_{i=1}^qH_it}=\lim_{n\rightarrow \infty }\left(
\prod_{i=1}^qe^{-iH_it/n}\right) ^n
\end{equation}
But, it is still quite difficult for a standard quantum computer
to simulate some highly correlated many body models. As an
example, let us consider ``coupled dimer''
Hamiltonian\cite{Gelfand}:
\begin{equation}
H_d=J\sum_{\left\langle i,j\right\rangle \in
~A}\overrightarrow{\sigma }_i \overrightarrow{\sigma }_j+\lambda
J\sum_{\left\langle i,j\right\rangle \in ~B}\overrightarrow{\sigma
}_i\overrightarrow{\sigma }_j
\end{equation}

\begin{figure}[tbh]
\epsfig{file=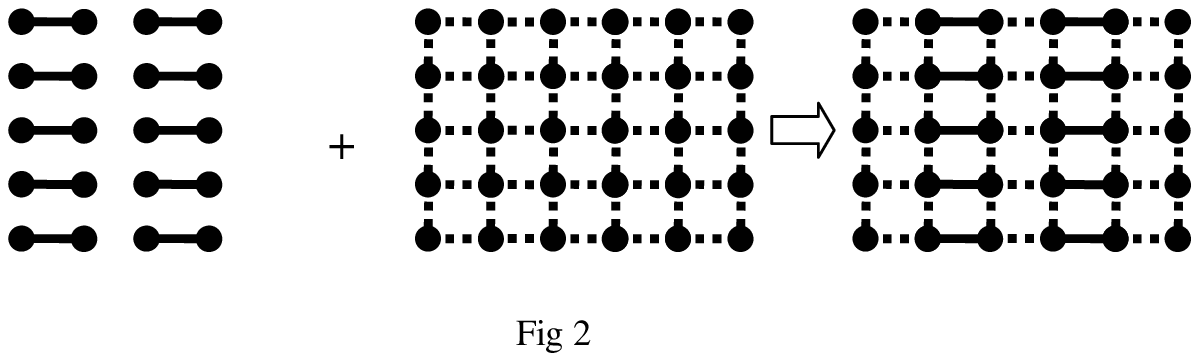,width=8cm} \caption{The coupled dimer
antiferromagnetic Hamiltonian $H_d$ is represented by the two
dimensional lattice on the right. The $A$ links are shown as thick
solid lines, and the $B$ links as thick dashed lines. $H_d$ can be
decomposed into the square lattice antiferromagnetic part and
decoupled dimer part. }
\end{figure}

Here, $A$ links form decoupled dimers while $B$ links couple the
dimers (see Fig.2). $\lambda $ is the dimensionless coupling
coefficient. To simulate the unitary evolution $\exp \{-iH_dt\}$
of this 2-dimensional lattice with $n^2$ sites, it will take
$O(n^{2K})$ elementary logical gate operations, if we partition
$\left[ 0,t\right] $ into $K$ identical intervals. However, the
overhead can be avoided if we use global operations. We may
decompose Hamiltonian $H_d$ into two parts: $H_d=H_1+H_2$, where
$H_1=\lambda J\sum_{\left\langle i,j\right\rangle \in A\cup
B}\overrightarrow{\sigma }_i\overrightarrow{\sigma }_j$ and
$H_2=\left( 1-\lambda \right) J\sum_{\left\langle i,j\right\rangle
\in A}\overrightarrow{\sigma }_i\overrightarrow{\sigma }_j$. $H_1$
is the square lattice anti-ferromagnetic Hamiltonian, which can be
directly realized by adjusting the potential of the optical
lattice. In addition, it is quite straightforward to implement
unitary transformations $U_a\left( \theta \right) =\exp \{i\theta
\sum_{\left\langle i,j\right\rangle \in A}\sigma _i^z\sigma
_j^z\}$  and $U_b\left( \theta \right) =\exp \{i\theta
\sum_{\left\langle i,j\right\rangle \in A}(\sigma _i^x\sigma
_j^x+\sigma _i^y\sigma _j^y)\}$ by using horizontal global
operations and parity unitary operations. Because $\sigma
_i^z\sigma _{j}^z$ commutes with $\sigma _i^x\sigma _{j}^x+\sigma
_i^y\sigma _{j}^y$, unitary transformation $\exp \{i\frac \theta
{(1-\lambda )J}H_2\}$ can be achieved by concatenating $U_a\left(
\theta \right)$ and $U_b\left( \theta \right)$. Hence, more
effective simulation for many body Hamiltonian $H_d$ can be
implemented by using global manipulation. By investigating this
example, we find it is very convenient to use global manipulations
to implement operations with periodic structures, although
focusing processes on single or two sites will take many iterative
steps. This suggests that quantum error-correction can be achieved
by taking advantage of global manipulations.

In addition, by adjusting the period and phase of the Hamiltonian
described by Eq. (3) we can implement the unitary operation $
\prod_{k=0}^{n/4}I_{4k}I_{4k+1}\sigma _{4k+2}^z\sigma _{4k+3}^z$.
Thus, an interval interaction form can be realized:
$\prod_{k=0}^{n/2}\exp (i\theta \sigma _{2k}^z\sigma _{2k+1}^z)$.
As shown by Ref\cite{Benjamin}, if one can realize four types of
Hamiltonian: $H_{2i}^s$, $H_{2i+1}^s$, $ H_{2i,2i+1}^{int}$, and
$H_{2i-1,2i}^{int}$ in one dimensional lattice, we can implement
quantum computation via cellular-automata approach. Here, $
H_{2i(2i+1)}^s$ and $H_{2i,2i+1(2i-1,2i)}^{int}$ refer to the
uniform single qubit operation on even (odd) sites and the
interaction between the even sites and its right (left) nearest
neighbor sites, separately.

\section{summary}

In this paper, we have devised a scheme for universal quantum
computation in the periodic lattice via global but periodic
operations. Our idea for quantum manipulation derive from
mathematical Fourier transformations, i.e., any a function with
space distribution can be decomposed into the superposition of
periodic eigen-functions. As a result, we derive the controls
localizing single site and two nearest neighboring sites by
adjusting global periodic Hamiltonians. We also show that
simulating nontrivial many body Hamiltonian and quantum
error-correcting operations will benefit from global
manipulations.

\emph{Note added:} After submission of this work, we noted two
schemes for implementing single-qubit operations in 2 dimensional
optical lattice\cite{Joo,Zhang}.

This work was supported by National Fundamental Research Program,
the Innovation funds from Chinese Academy of Sciences, National
Natural Science Foundation of China (Grant No. 10574126,
60121503), NCET-04-0587, CPSF(2005038012), and CAS K. C. Wong
Post-doctoral Fellowships, Z-W. Zhou thanks L.-M. Duan, L.-A. Wu,
H. Pu and X. Zhou for helpful discussions.

\end{document}